\documentclass[twocolumn,prl]{revtex4}

\usepackage{graphicx}
\usepackage{dcolumn}
\usepackage{bm}
\usepackage{multirow}
\usepackage{color}
\usepackage[normalem]{ulem}
\usepackage{amsmath}
\usepackage{gensymb}
\usepackage[colorlinks=true]{hyperref}

\begin{document}

\title{Exciton spectroscopy and diffusion in MoSe$_2$-WSe$_2$ lateral heterostructures encapsulated in hexagonal boron nitride}

\author{Dorian Beret$^{1,+}$}
\author{Ioannis Paradisanos$^{1,+}$}
\author{Ziyang Gan$^{2,+}$}
\author{Emad Najafidehaghani$^2$}
\author{Antony George$^{2,3}$}
\author{Tibor Lehnert$^{4 \ddagger}$}
\author{Johannes Biskupek$^4$}
\author{Shivangi Shree$^5$}
\author{Ana Estrada-Real$^1$}
\author{Delphine Lagarde$^1$}
\author{Jean-Marie Poumirol$^6$}
\author{Vincent Paillard$^6$}
\author{Kenji Watanabe$^7$}
\author{Takashi Taniguchi$^8$}
\author{Xavier Marie$^1$}
\author{Ute Kaiser$^4$}
\author{Pierre Renucci$^1$}
\author{Laurent Lombez$^1$}
\email{laurent.lombez@cnrs.fr}
\author{Andrey Turchanin$^{2,3}$}
\email{andrey.turchanin@uni-jena.de}
\author{Bernhard Urbaszek$^1$}
\email{physik.bu@gmail.com}

\affiliation{\small$^1$Universit\'e de Toulouse, INSA-CNRS-UPS, LPCNO, 135 Avenue Rangueil, 31077 Toulouse, France}
\affiliation{\small$^2$Friedrich Schiller University Jena, Institute of Physical Chemistry, 07743 Jena, Germany}
\affiliation{\small$^3$Abbe Centre of Photonics, 07745 Jena, Germany}
\affiliation{\small$^4$Ulm University, Central Facility of Electron Microscopy, D-89081 Ulm, Germany}
\affiliation{\small$^5$Department of Physics, University of Washington, Seattle, WA, USA.}
\affiliation{\small$^6$CEMES-CNRS, Universit\'e de Toulouse, Toulouse, France}
\affiliation{\small$^7$Research Center for Functional Materials, National Institute for Materials Science, 1-1 Namiki, Tsukuba 305-0044, Japan}
\affiliation{\small$^8$International Center for Materials Nanoarchitectonics, National Institute for Materials Science, 1-1 Namiki, Tsukuba 305-0044, Japan}

\begin{abstract}
Chemical vapor deposition (CVD) allows lateral edge epitaxy of  transition metal dichalcogenide heterostructures with potential applications in optoelectronics. 
Critical for carrier and exciton transport is the quality of the two materials that constitute the monolayer and the nature of the lateral heterojunction. Important details of the optical properties were inaccessible in as-grown heterostructure samples due to large inhomogeneous broadening of the optical transitions. 
Here we perform optical spectroscopy at T = 4~K and also at 300~K to access the optical transitions in CVD grown MoSe$_2$-WSe$_2$ lateral heterostructures that are transferred from the growth-substrate and are encapsulated in hBN. Photoluminescence (PL), reflectance contrast and Raman spectroscopy reveal considerably narrowed optical transition linewidth  similar to high quality exfoliated monolayers. In high-resolution transmission electron microscopy (HRTEM) we find near-atomically sharp junctions with a typical extent of 3~nm for the covalently bonded MoSe$_2$-WSe$_2$. 
In PL imaging experiments we find effective excitonic diffusion length that are longer for WSe$_2$ than for MoSe$_2$ at low T=4~K, whereas at 300 K this trend is reversed.

\end{abstract}

\maketitle

\textbf{Introduction.---} 
Research on semiconducting transition metal dichalcogenide (TMDs) monolayers \cite{Splendiani:2010a,Mak:2010a,tonndorf2013photoluminescence} and their heterostructures is motivated by new collective effects of the electronic system \cite{kennes2021moire,gu2022dipolar,li2021imaging} and the potential for new optoelectronic and quantum technology devices \cite{wang2018electroluminescent,Novoselov:2016a,Mak:2016a,Schaibley:2016a,unuchek2018room,Schneider2018a,Koperski:2017a,dufferwiel2017valley,Scuri:2018a,hong2014ultrafast,quantumgerardot2022}.
The optical and electronic properties of these materials can be tuned by combining different monolayers (MLs) via van der Waals stacking to create vertical heterostructures \cite{geim2013van,shree2021guide,andrei2021marvels}. But interestingly, tuning of optical properties of TMDs can also be achieved while staying in the ultimate monolayer limit. Recent progress is based on innovative growth techniques such as for
Janus monolayers with different top and bottom chalcogen \cite{lu2017janus,zheng2021excitonic,petric2021raman} as well as in lateral heterostructures (LHs) \cite{najafidehaghani20211d,huang2014lateral} within the monolayer plane with an atomically-sharp 1D interface which exhibits p-n junction characteristics \cite{duan2014lateral,gong2014vertical}. 
Examples of potential applications for LHs are
photodetectors \cite{wu2018self}, p–n junction diodes \cite{najafidehaghani20211d,pospischil2014solar,sahoo2018one}, photovoltaic \cite{pospischil2014solar}, electroluminescent \cite{pospischil2014solar} and quantum devices \cite{wang2019recent}. 
\\
\begin{figure*}
\includegraphics[width=0.98\linewidth]{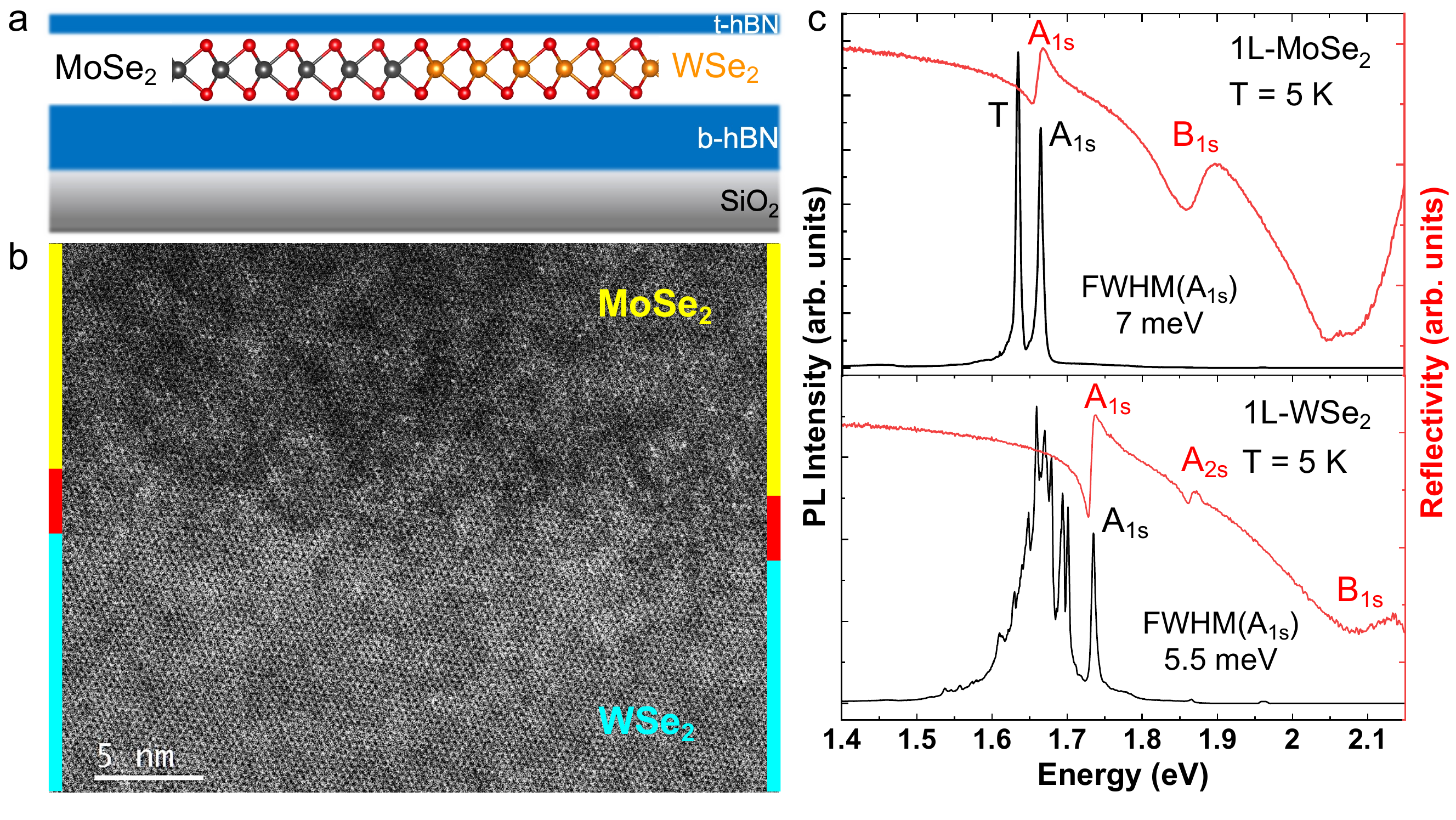}
\caption{\label{fig:fig1} \textbf{Lateral heterostructure spectroscopy and microscopy.} (a) Schematic representation of the MoSe$_2$-WSe$_2$ LH, encapsulated in hBN. Grey, orange and red spheres represent Mo, W and Se atoms, respectively. (b) HAADF-STEM image showing a MoSe$_2$-WSe$_2$ boundary (between the red lines). The MoSe$_2$ and WSe$_2$ areas are located between the yellow and the cyan lines, respectively. (c) Corresponding PL (black) and reflectance contrast (red) spectra of MoSe$_2$ (top) and WSe$_2$ (bottom). The PL linewidth of neutral excitons (A$_{1s}$) is 7 meV in MoSe$_2$ and 5.5 meV in WSe$_2$.}
\end{figure*}
\indent LHs can be fabricated following one-step \cite{sahoo2018one,huang2014lateral,gong2014vertical,najafidehaghani20211d} or multiple-step \cite{li2015epitaxial,zhang2018edge} growth processes either by physical vapor deposition (PVD) or by chemical vapor deposition (CVD). Electron beam lithography has also been used for the fabrication of LHs \cite{mahjouri2015patterned}. One-step CVD approaches with suitable growth conditions are simpler and have the advantage to grow large area TMD LHs at lower temperatures \cite{najafidehaghani20211d}. Accessing the quality of the monolayer heterojunction is so far mainly based on electron microscopy techniques. A depletion width on the order of few nanometers is reported using scanning tunneling microscopy and spectroscopy techniques \cite{chu2018atomic}.  The electronic structures and band alignments of TMD LHs have been calculated using density functional theory \cite{guo2016band} and the formation of interface excitons is predicted by tight-binding models, as well as effective mass models \cite{lau2018interface}. \\
\indent To further access carrier dynamics and excitonic properties at the interface optical spectroscopy is needed as a powerful and non-invasive tool. 
But in as-grown CVD samples details are masked due to the large inhomogeneous broadening reported for the optical transitions. The vast majority of optical characterization experiments performed on LHs are reported at room temperature and a detailed investigation at cryogenic temperatures is not reported so far. At cryogenic temperatures, a precise estimation of the material quality near and across the junction are possible, as thermal broadening effects are reduced and defects can be detected through luminescence of localized states. 
An important target is investigating the consequences of connecting MoSe$_2$, a material dominated by bright spin-allowed exciton transitions \cite{lu2019magnetic,robert2020measurement}, with WSe$_2$ which has a very similar bandgap but shows very rich and striking emission features based on dark, spin-forbidden exciton complexes \cite{PhysRevB.105.085302}.\\
\begin{figure*}
\includegraphics[width=0.98\linewidth]{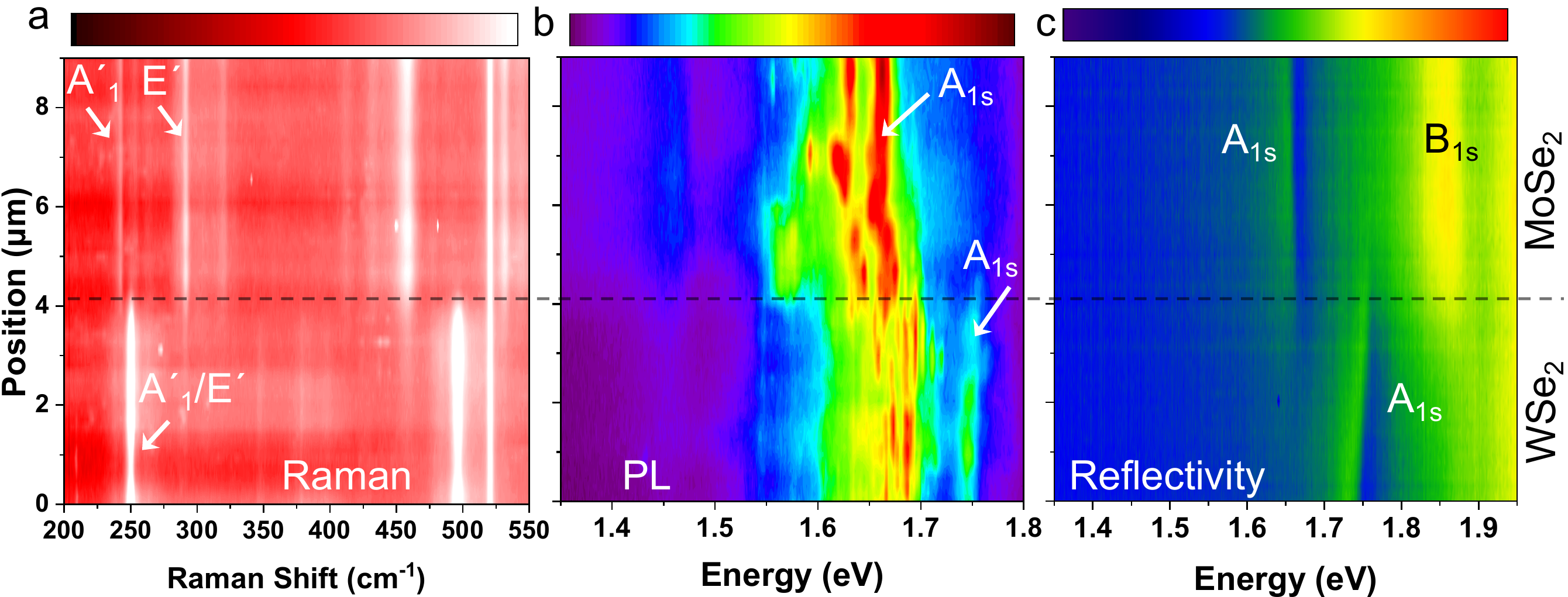}
\caption{\label{fig:fig2} \textbf{Lateral heterostructure optical spectroscopy scans.}  Contour plot of (a) Raman ($\lambda_L=633~$nm), (b) PL ($\lambda_L=633~$nm) and (c) reflectivity spectra scans across a WSe$_2$-MoSe$_2$ heterojunction at T=5~K. The black horizontal dashed line indicates the position of the heterojunction. Color bars for panels (a), (b) and (c) are normalized from 0 to 1.}
\end{figure*}
\indent Here we perform low-temperature optical spectroscopy and microscopy experiments in CVD-grown MoSe$_2$-WSe$_2$ monolayer LHs \cite{najafidehaghani20211d}. We first lift the LHs from the growth substrate. This shows that transfer of the lateral heterostructure to other substrates is possible for device processing. Second, we encapsulate the LHs in high quality exfoliated hBN flakes\cite{Taniguchi:2007a} (Fig.~\ref{fig:fig1}a). Encapsulation of the TMD monolayer in high quality hexagonal boron nitride (hBN) \cite{Taniguchi:2007a} is crucial to access the intrinsic optical properties of exfoliated and CVD grown flakes \cite{rhodes2019disorder,Archana2019a,Cadiz:2017a,ajayi2017approaching,wierzbowski2017direct,Stier:2018a,shree2019high,paradisanos2020controlling,2018arXiv181009834M}. 
We report optical transition linewidth of the LH monolayer ($\approx 5~$meV at T=4~K) comparable to high quality exfoliated layers.
Our step-by-step scans across the heterojunctions in optical spectroscopy experiments show an abrupt change from WSe$_2$ to MoSe$_2$. In atomic-resolution transmission electron microscopy on our samples we find a transition with a nm-sharp junction from MoSe$_2$ to WSe$_2$. The structural quality is  also demonstrated in Raman spectroscopy. Photoluminescence (PL) imaging experiments allow us to investigate excitonic transport  governed by the different effective lifetime of the exciton species for MoSe$_2$ and  WSe$_2$ at T = 4~K and 300~K. \\

\begin{figure*}
\includegraphics[width=0.8\linewidth]{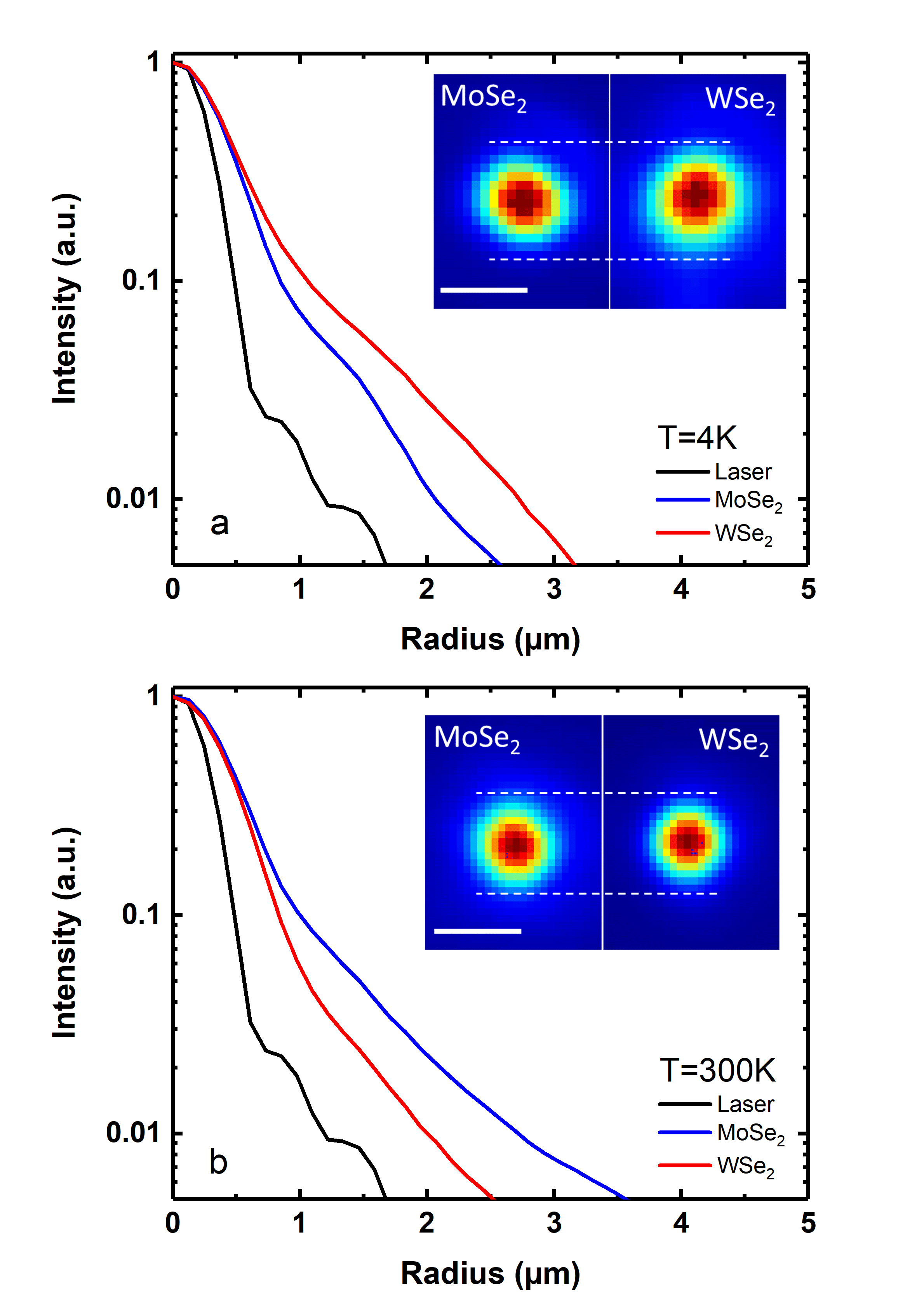}
\caption{\label{fig:fig3} \textbf{Photoluminescence imaging.} (a) Exciton diffusion measurements at a temperature of T=4~K for the MoSe$_2$ monolayer region (blue) and WSe$_2$(red). In the inset we show the MoSe$_2$ diffusion spot compared to the WSe$_2$ spot. The laser spot diameter (black) is shown to indicate over which area excitons are initially generated, with small intensity oscillations due to Airy discs visible in logarithmic scale (b) same as (a) but at T = 300~K.}
\end{figure*}

\begin{figure*}
\includegraphics[width=0.7\linewidth]{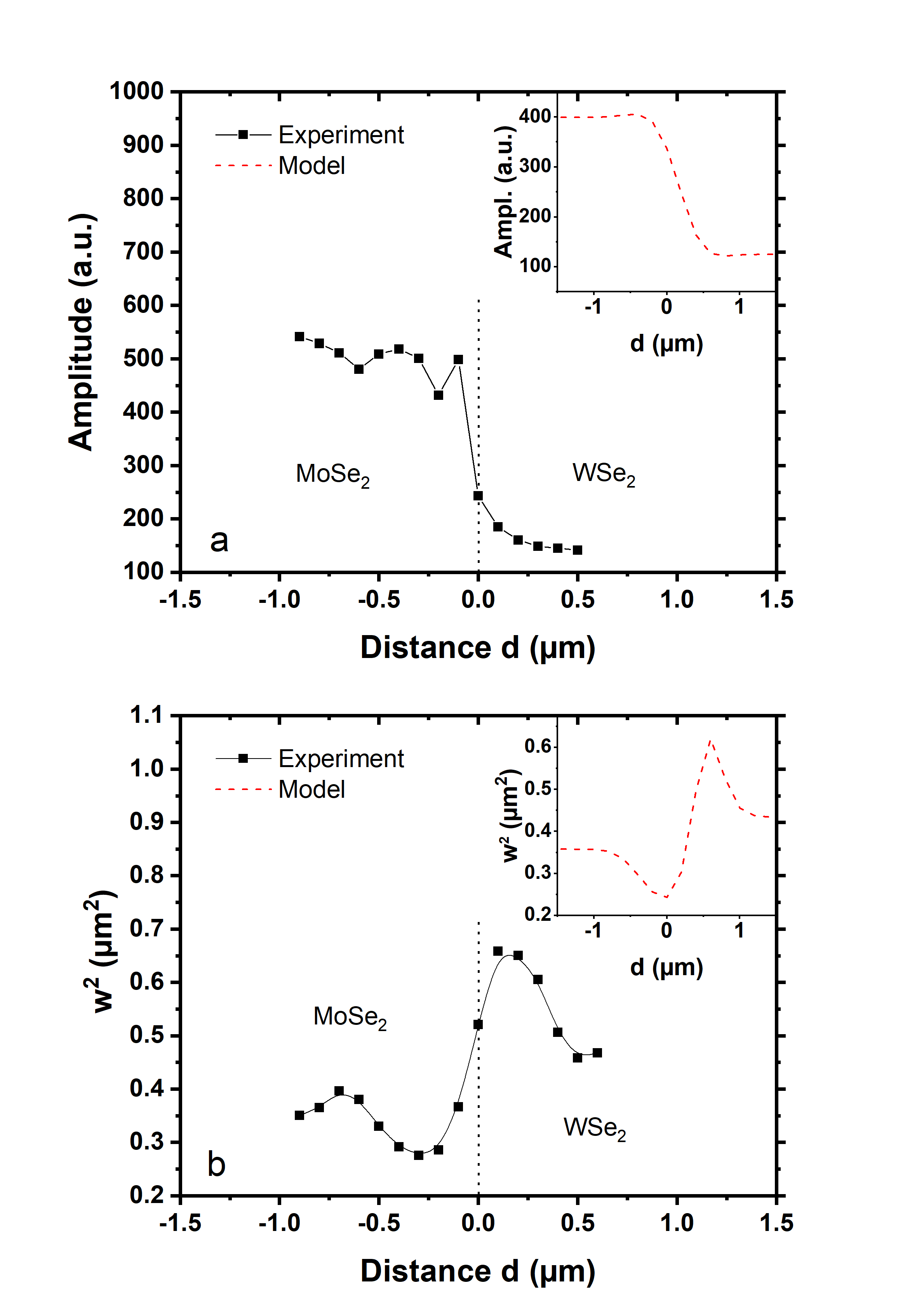}
\caption{\label{fig:fig4} \textbf{Photoluminescence imaging scans at T= 4~K.} (a) Amplitude of the PL profile when crossing the lateral interface (b) Squared width of the PL profile when crossing the lateral interface. The insets show results from a simple numerical model that captures the main features.}
\end{figure*}

\textbf{Experimental results.---}
Our MoSe$_2$-WSe$_2$ lateral monolayer heterojunction is grown by CVD synthesis that we reported recently \cite{najafidehaghani20211d}.
A schematic representation of the encapsulated MoSe$_2$-WSe$_2$ structure is shown in Fig.~\ref{fig:fig1}a. Fig.~\ref{fig:fig1}b shows high-angle annular dark field scanning transmission electron microscopy (HAADF-STEM) image recorded at  the boundary region of the MoSe$_2$-WSe$_2$  structures. The image was recorded at an operating voltage of 200~kV, see supplement. The atomically resolved HAADF image allows the detection of the roughness of the interface at sub-nm resolution due to exploiting of the Z-contrast, where lighter Mo atoms show darker contrast than heavier W atoms \cite{liu2005scanning}. Thus, MoSe$_2$ appears darker and WSe$_2$ appears brighter in contrast. From these measurements we estimate the transition width of the boundary region between MoSe$_2$ and WSe$_2$ to be as narrow as $\approx$3~nm.\\
\indent We use water-assisted deterministic transfer to pick up as-grown, CVD LHs using polydimethylsiloxane (PDMS) and deterministically transfer and encapsulate them in hBN \cite{jia2016large,paradisanos2020controlling}. We use these encapsulated samples for all optical spectroscopy measurements. \\
\indent We first analyze the PL and differential white light reflectivity spectra of the individual monolayer areas (away from the heterojunction), collected at T=5~K. Fig.~\ref{fig:fig1}c shows superimposed PL (black) and reflectivity (red) spectra of MoSe$_2$ (top) and WSe$_2$ (bottom) monolayers after transfer and encapsulation in hBN. 
We measure in PL a neutral exciton (A$ _{1s} $) linewidth of 7 meV in MoSe$_2$ and 5.5 meV in WSe$_2$ (Fig.~\ref{fig:fig1}c). These FWHM values are comparable to high-quality exfoliated MoSe$_2$ and WSe$_2$ MLs \cite{ajayi2017approaching,wierzbowski2017direct,Stier:2018a,Scuri:2018a}. We note that homogeneous and inhomogeneous broadening effects contribute to the total optical transition linewidth of any TMD ML \cite{shree2021guide}. The homogeneous part is defined by the radiative lifetime and is of the order of 1~meV. Thus, any substantially larger linewidth at low temperature will provide the inhomogeneous contribution due to imperfections in the ML or the direct environment \cite{Moody:2015a}. MoSe$_2$ monolayers exhibit two pronounced PL emission peaks at 1.665~eV and 1.635~eV, assigned to neutral neutral (A$ _{1s} $) and charged (T) excitons, respectively \cite{shree2018observation}. In the case of WSe$_2$, the A$ _{1s} $ transition is located at 1.734~eV with the singlet trions (T) and dark excitons (X$^{D}$) lying 35~meV and 40~meV below A$ _{1s} $, consistent with previous reports \cite{wang2017plane}. Additional peaks appear at lower energies, attributed to contributions from neutral, charged biexcitons and localized emission from defects. In Fig.~\ref{fig:fig1}c strong excitonic resonances appear in differential reflectivity for both materials with negligible Stokes shift in energy compared to the PL emission, which is a sign of negligible exciton localization effects for these spectra. Clear signatures of B$ _{1s} $ exciton states are also observed in both materials. For WSe$_2$ the appearance of the A$ _{2s} $ excited exciton state, which has a larger Bohr radius, shows the good quality of the CVD-grown monolayers of the LHs \cite{delhomme2019magneto}.\\
\indent To optically examine the interface between the two materials, we perform linescans across several heterojunctions, see supplement. In particular, we collect the Raman, PL and reflectivity spectra within our detection spot while moving the sample over a $ \approx $~10$ ~\mu $m distance with a step size of $ \approx $~150~nm using attocube nanopositioners. Typical Raman PL and reflectivity scans across a heterojunction are shown in the contour plots of Fig.~\ref{fig:fig2}a,b,c. A black dashed line indicates the position of the heterojunction in each contour plot. 
As a common feature for all three contour plots, we observe abrupt changes in the optical spectra as we scan across the lateral heterojunction, as a result of distinctly different phonon energies and exciton transition energies in the two materials.\\ 
\indent Now we discuss Raman spectroscopy results in Fig.~\ref{fig:fig2}a for ML MoSe$_2$ and WSe$_2$, collected at T=4~K using an excitation laser wavelength of $ \lambda $=633~nm. The main Raman peaks of MoSe$_2$ and WSe$_2$ can be identified in the contour plot (detailed Raman spectra can be found in the Supplementary Material, section B). For ML MoSe$_2$ we observe the A$'_{1}$($\Gamma$) phonon at 241~cm$^{-1}$ and the E$'$($\Gamma$) at 291~cm$^{-1}$ \cite{soubelet2016resonance}, while a strong peak at 458~cm$^{-1}$ has been associated to other peaks to form triplets \cite{mcdonnell2020observation}. Interestingly, we also observe a strong peak at 531~cm$^{-1}$, recently assigned to multi-phonon processes associated with either both K and M point phonons or a combination of $\Gamma$ point phonons \cite{mcdonnell2020observation}. The observation of this Raman peak is a signature of resonant excitation with an excited exciton state in MoSe$_2$. WSe$_2$ phonons are spectrally different compared to MoSe$_2$. The degenerate A$'_{1}$($\Gamma$)/E$'$($\Gamma$) phonons are located at 250~cm$^{-1}$ and, similar to MoSe$_2$, a strong and recently discovered peak at 495~cm$^{-1}$ is also observed here and attributed to multi-phonon processes at K and M points or combination of $\Gamma$ point phonons \cite{mcdonnell2020observation}. The Raman analysis and the assignment of the Raman peaks further confirms high quality CVD-grown MLs. As the A$'_{1}$($\Gamma$) and E$'$($\Gamma$) modes remain practically constant in the Raman linescans (Fig.~\ref{fig:fig2}a), we tentatively infer that there are not any significant variations in the electron density \cite{sohier2019enhanced} and lattice structure \cite{dadgar2018strain} close to the heterojunction. \\
\indent The PL linescan in Fig.~\ref{fig:fig2}b shows emission from the main exciton transitions in WSe$_2$ and MoSe$_2$ (indicated by arrows) and a clear change in transition energy is discernible as we go across the lateral junction, with similar energies as for the individual spectra shown in Fig.~\ref{fig:fig1}c. Due to the extreme sensitivity of the PL emission energy spectrum on the defect concentration and dielectric environment \cite{shree2021guide}, we see spectral shifts between spectra taken at different positions. This makes the PL contour plot appear less smooth than the Raman and reflectivity linescans. \\
\indent The reflectivity linescan shown in Fig.~\ref{fig:fig2}c shows a clear change in the main exciton energies as we scan from one material to the other. The main neutral exciton A$_{1s}$ feature in MoSe$_2$ in reflectivity is at an energy of 1.665~eV, whereas for WSe$_2$ this transition energy is clearly shifted to 1.734~eV. Reflectivity is less sensitive to the local defect and dielectric environment and therefore the measured transition energies remain comparatively constant from one spectrum to the other.\\

\textbf{Exciton diffusion.---} A PL imaging experiment is used to probe the effective diffusion length in the two TMD materials that are connected at the junction. First, we concentrate on results away from the lateral junction i.e. neither the excitation nor the PL emission profile have any spatial overlap with the lateral junction. The experiment consists in a local photogeneration of excitons with a He:Ne laser focused on a spot size of about 0.7~$\mu$m (see black curve in Fig.~ \ref{fig:fig3}) and with an excitation power of 5~$\mu$W. The generated spatial gradient of the exciton concentration (i.e. chemical potential gradient) induces lateral diffusion of excitons which is probed by recording the spatial PL profile with a CMOS camera, see supplement, where we intergate over the full spectral range of the monolayer emission. Fig.~\ref{fig:fig3}a shows the PL profiles obtained at T=4~K on WSe$_2$ (red line) and MoSe$_2$ (blue line). We clearly observe that PL emission occurs over a larger spot diameter than the laser excitation \cite{cadiz2018exciton,PhysRevLett.120.207401}. At this temperature WSe$_2$ shows a longer effective diffusion length than MoSe$_2$. This difference is also visible in the inset by directly comparing the images of the PL profile from the two materials. \\
\indent Interestingly, this behavior is reversed at T=300~K where MoSe$_2$ shows a longer effective diffusion length than WSe$_2$ (see Fig.~\ref{fig:fig3}b). This is possibly linked to dark excitonic states which have longer lifetimes (PL emission times) \cite{zhang2015experimental,robert2017fine}. Indeed, the lowest energy state of the conduction band is a dark state in WSe$_2$ while it is a bright state in MoSe$_2$. It has therefore been reported that their PL intensity evolution with the temperature shows opposite trends: WSe$_2$ is darker at low temperature and brighter at room temperature as compared to MoSe$_2$ \cite{wang2015spin}. The contribution of dark exciton states with longer lifetime would increase the effective diffusion length as compared to MoSe$_2$, as dark exciton emission is negligible for MoSe$_2$ \cite{lu2019magnetic,robert2020measurement}. Therefore, when we compare the two materials WSe$_2$ has a longer diffusion length at 4~K and a shorter one at T=300~K.\\
\indent We then proceed to another experiment where we scan the laser excitation spot across the lateral WSe$_2$-MoSe$_2$ junction and record a PL image at each location, with a step size of about 100~nm between each acquisition. For the measurements in Fig.~\ref{fig:fig4} we have chosen an interface region which gives uniform PL images, for more statistics on PL images see supplement. For our scan across the lateral interface, we fit for each image the PL profile by a Gaussian distribution from which we extract the amplitude and the width w$^2$. The evolution of these two parameters is displayed in Fig.~\ref{fig:fig4} for the measurement at T=4~K. The amplitude (proportional to the PL intensity) decreases as we go from MoSe$_2$ to WSe$_2$. The width w$^2$ of the PL profile shows a more complex behavior, similar to an oscillation. We thus performed a modeling of the experiment by solving a two-dimensional diffusion equation in the two materials with a finite element method, see supplement. The modeling results are displayed in the insets in Fig.~\ref{fig:fig4}. They show qualitative agreement with the experimental data and indicate a smooth transition from one material to the other, in agreement with our TEM microscopy data in Fig.~\ref{fig:fig1}b. 
 Our basic model reproduces the data when we consider as the only input information that WSe$_2$ has a longer effective exciton lifetime and a lower radiative efficiency than MoSe$_2$ at T=4~K. This scenario is consistent with the involvement of dark states for WSe$_2$ emission and might lie at the origin of the observed change in width w$^2$ of the PL profile as a function of scan distance.\\
 
\indent \textbf{In conclusion}, we have performed detailed spectroscopic studies on CVD grown lateral MoSe$_2$-WSe$_2$ heterostructures. As an important step towards device processing and for accessing the intrinsic optical quality of the junction, we have first transferred the sample from its original growth substrate and then encapsulated the lateral heterostructures in top and bottom flakes of high quality hBN.
Our experiments give access to the excitonic structures at cryogenic temperatures, with neutral exciton transition linewidth of the order of 5 meV. In exciton diffusion experiments we show that the MoSe$_2$ and WeSe$_2$ exciton transport show opposite trends in temperature dependent experiments, as dark excitons contribute to the PL signal in WSe$_2$ and not in MoSe$_2$. In future studies on these high-quality samples a clearer identification of interlayer excitons at the lateral junction is the target. Here an important consideration is the large spot diameter (order of 500 nm) of the excitation and detection spot in PL, whereas the lateral heterostructure border is only a 1-3~nm wide stripe (as we see in TEM) within this spot. So to increase the sensitivity for detecting photons originating from the junction region, it would be desirable to perform optical spectroscopy with higher spatial resolution. One promising direction here is to perform tip enhanced PL \cite{LeeLeeKangKooKimPark+2020+3089+3110,darlington2020imaging,zhang2022nano}. \\

\section{Supplement}

\subsection{Optical microscope image and linescan description} \label{partA}

\begin{figure}
\includegraphics[width=0.95\linewidth]{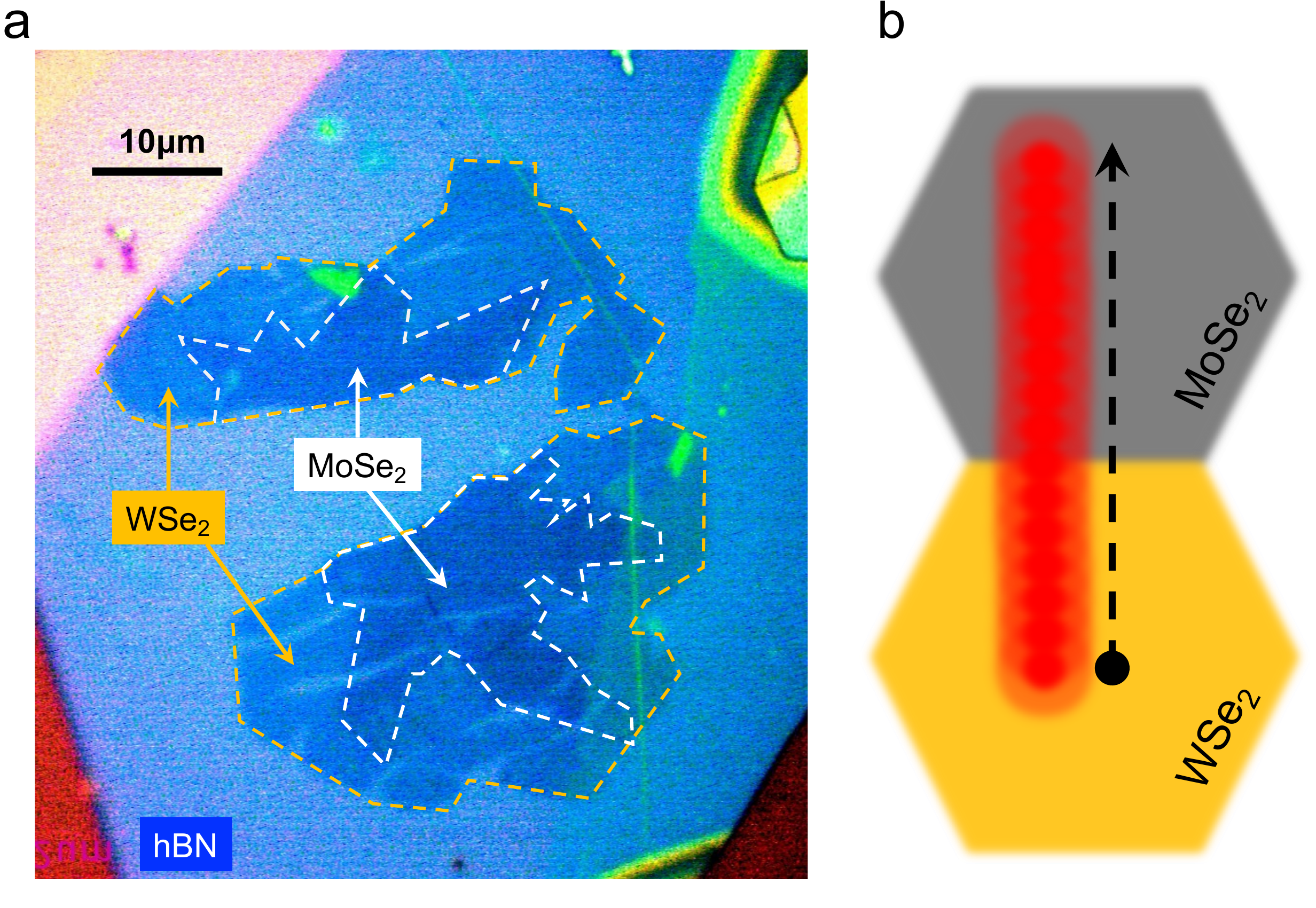}
\caption{\label{fig:figS1} (a) Optical image of the hBN-encapsulated LH. Orange and white dashed lines indicate the boundaries of WSe$_2$ and MoSe$_2$, respectively. (b) Schematic representation of the Raman, PL and Reflectivity linescan process across the heterojunction of MoSe$_2$/WSe$_2$ LH. The laser is initially positioned on WSe$ _2 $ and a series of sequential steps (step size $ \approx150$ nm) follows with a single signal acquisition after each step.}
\end{figure}

An optical microscope image of the MoSe$_2$/WSe$_2$ lateral heterostructure (LH) and a schematic representation of the linescan process used in the optical spectroscopy experiments are shown in Fig.~\ref{fig:figS1}a,b. We use water-assisted deterministic transfer to pick up as-grown, chemical vapor deposition (CVD) LHs using polydimethylsiloxane (PDMS) and deterministically transfer and encapsulate them in hBN \cite{jia2016large,paradisanos2020controlling}. The optical contrast between MoSe$_2$ and WSe$_2$ is different, allowing the direct visualization of the heterojunctions, shown with white dashed lines in Fig.~\ref{fig:figS1}a. In Fig.~\ref{fig:figS1}b we show a schematic representation of the process for the Raman, PL and Reflectivity linescans (Figure 2 in the main text). We position the excitation laser ($ \lambda = 633 $~nm, diffraction-limited spot diameter of 1~$ \mu $m) on the area of WSe$_2$ and we use piezo-controlled nanopositioners to move the sample with steps of $\approx150$~nm. After each step, a single -Raman, photoluminescence, reflectivity- spectrum is collected and finally all spectra are plotted as contour representations in Figure 2 of the main text.\\
\subsection{Optical spectroscopy}
\begin{figure*}
\includegraphics[width=0.9\linewidth]{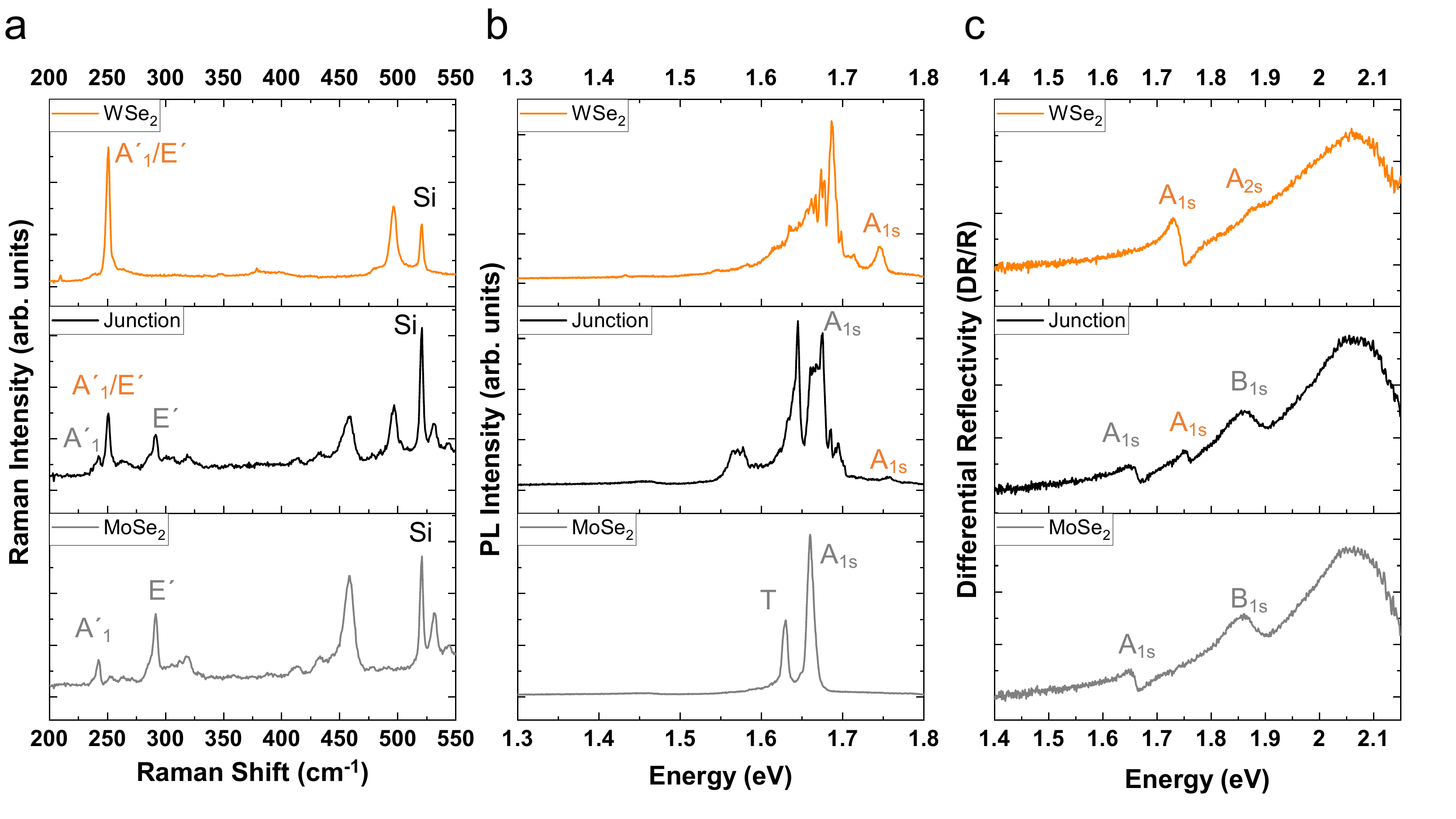}
\caption{\label{fig:figS2} (a) Raman, (b) photoluminescence and (c) reflectivity spectra of MoSe$_2$ (bottom), on the junction (middle) and WSe$_2$ (top).}
\end{figure*}
Raman, PL and differential white light reflectivity spectra are collected at T=5 K in a closed-loop liquid helium (LHe) system. Fig.~\ref{fig:figS2}a,b,c shows individual Raman, PL and reflectivity spectra of MoSe$_2$ (bottom), junction (middle) and WSe$_2$ (top), after transfer and encapsulation in hBN. For the Raman and PL experiments we use a 633~nm HeNe laser as an excitation source with a spot size diameter of $ \approx 1~\mu $m and 6~$ \mu $W power.  In reflectivity we use a tungsten-halogen white light source with a power of a few nW to collect the intensity reflection coefficient of the sample with the monolayer ($R_{ML}$) and the reflection coefficient of the substrate ($R_{S}$) so that $\Delta R=(R_{ML}-R_{S})/R_{S}$. 
The spectral shape and amplitude of the differential reflectivity signal also depends on thin-film interference effects related to the thickness of top, bottom hBN and SiO$_2$ layers \cite{Robert:2018a}\\
The main Raman peaks of MoSe$_2$ (Fig.~\ref{fig:figS2}a) include the A$'_{1}$($\Gamma$) phonon at 241~cm$^{-1}$ and the E$'$($\Gamma$) at 291~cm$^{-1}$ \cite{soubelet2016resonance}, while a strong peak at 458~cm$^{-1}$ has been associated two other peaks to form triplets, see supplementary material of \cite{mcdonnell2020observation}. A strong peak at 531~cm$^{-1}$ was recently assigned to multi-phonon processes either associated with both K and M point phonons or combinations of $\Gamma$ point phonons \cite{mcdonnell2020observation}. The observation of this phonon is a signature of resonant excitation with an excited exciton state. WSe$_2$ phonons are spectrally different compared to MoSe$_2$. The degenerate A$'_{1}$($\Gamma$)/E$'$($\Gamma$) phonons are located at 250~cm$^{-1}$ and, similar to MoSe$_2$, a strong and recently discovered peak at 495~cm$^{-1}$ is also observed here and attributed to multi-phonon processes at K and M points or combination of $\Gamma$ point phonons \cite{mcdonnell2020observation}. As expected, when the laser spot is on the junction, Raman spectra display a superposition of the individual spectral signatures. We did not identify any measurable shift or broadening of the Raman peaks when we scanned across the junction. PL spectra of MoSe$_2$ and WSe$_2$ monolayers exhibit pronounced peaks (Fig.~\ref{fig:figS2}b), associated with neutral neutral (X$^{0}$) and charged (X$^{T}$) excitons. Additional peaks appearing at lower energies, especially in WSe$_2$ and at the junction, possibly due to an ensemble of localized emission from defects. Further experiments are necessary to examine any possibility of the formation of interlayer excitons. Strong excitonic resonances appear in reflectivity for both materials with negligible Stokes shift between emission and absorption of the A1s exciton state (see Fig.~\ref{fig:figS2}c). Clear signatures of B1s states are also observed in both materials while for WSe$_2$ the appearance of the A2s excited state further supports the good quality of the CVD-grown monolayers. Similar to Raman and PL spectroscopy, on the junction we observe a superposition of the available states from both materials.\\

\subsection{Diffusion experiment and model}
For the investigation of lateral transport, we used an experimental procedure very similar to the one described in part \ref{partA}. The excitation is based on a HeNe laser with an excitation spot size of about 0.7$\mu$m and an excitation power of 5$\mu$W. The PL images are recorded by a Hamamatsu Fusion-BT CMOS camera. The PL profiles are plotted in polar coordinates. To do so, we integrate over all the polar angles [0;2$\pi$], the data are therefore $I_{PL}(r)$ where $r$ is the radius from the maximum PL intensity. \\
\indent To model the scanning part over the junction (see Fig.4 in the main text), we numerically solved the 2D classical diffusion equation with the use of the Matlab toolbox \textit{pdetool} (partial differential equation toolbox). We set two zones having the same diffusion coefficient of $1cm^2/s$ but different lifetime of $\tau_1=100ps$ and $\tau_2=400ps$. To take into account the contribution of dark and bright states, different radiative efficiency $\eta$ are taken into account with a ratio of $\eta_1/\eta_2=10$. The laser excitation spot is assumed gaussian, and we model the PL profile for each position of the laser spot. The PL profiles $I_{PL}(r)$ are fitted by a gaussian distribution $I_{PL}(r)=A.exp(-r^2/w^2)$ where $w^2$ reflects the broadening of the PL profile (i.e. diffusion). We used Neumann boundary conditions with $dI_{PL}(r)/dr=0$ to simulate ideal interface with no recombination.
\\
\subsection{Electron microscopy}

For STEM investigation, the samples were transferred to Quantifoil$^{tm}$ grids  using PMMA assisted transfer protocol. The High-angle annular dark-field scanning transmission electron microscopy (HAADF-STEM) image was acquired with a Thermofisher Talos 200X microscope operated at 200 kV. 

\begin{figure*}
\includegraphics[width=0.8\linewidth]{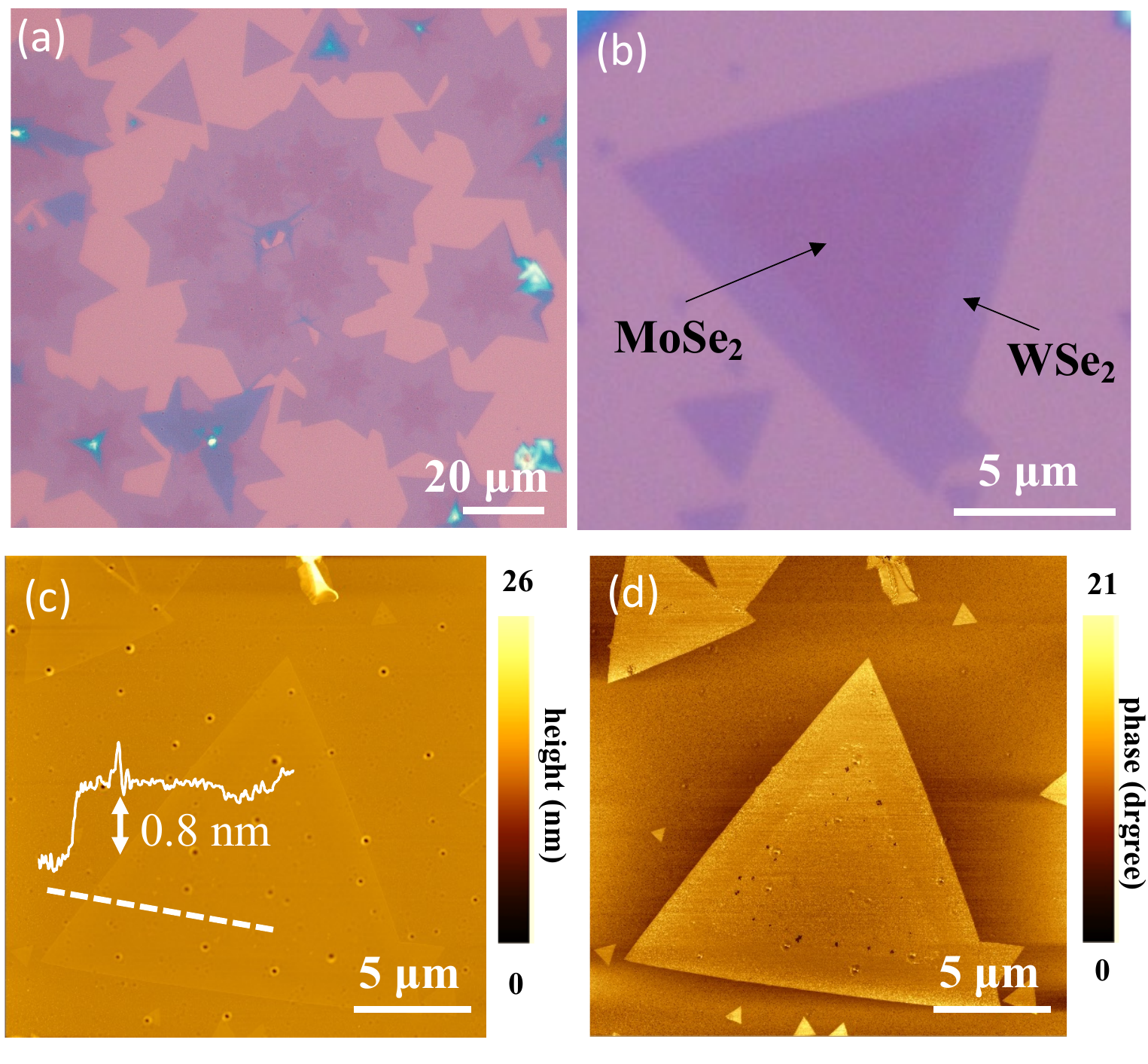}
\caption{\label{fig:figS3} \textbf{Characterization of lateral heterostructures }(a,b) Optical microscopy image of as grown MoSe$_2$-WSe$_2$ LHs (c,d) AFM height and phase image of as grown LHs. The height profile in the inset of (c) shows monolayer thickness of 0.8~nm. }
\end{figure*}

\begin{figure*}
\includegraphics[width=0.8\linewidth]{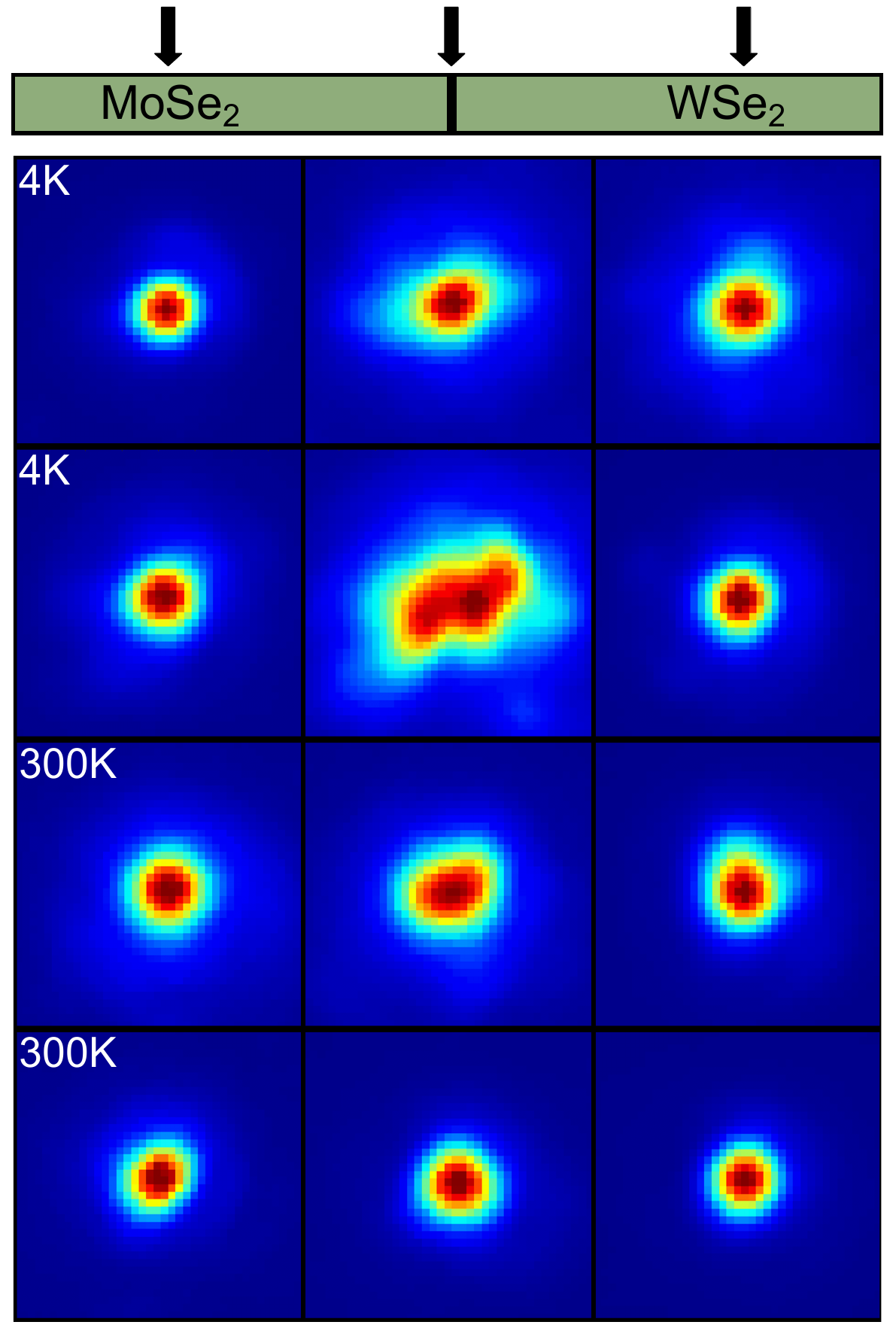}
\caption{\label{fig:figS4} \textbf{Examples of PL imaging in different sample areas}. We show more examples of typical PL emission spots in MoSe$_2$ (left), on the junction between the two materials (middle), and in WSe$_2$ (right) for two different sample temperatures as marked on the panels for each line.}
\end{figure*}

 \textbf{Acknowledgements}  \\
Toulouse acknowledges  partial funding from ANR IXTASE, ANR HiLight, NanoX project 2DLight, the Institut Universitaire de France, and the EUR grant ATRAP-2D NanoX ANR-17-EURE-0009 in the
framework of the "Programme des Investissements d'Avenir", the Institute of quantum technology in Occitanie IQO and a UPS excellence PhD grant.
Growth of hexagonal boron nitride crystals was supported by JSPS KAKENHI (Grants No. 19H05790, No. 20H00354 and No. 21H05233). 
The Jena group received financial support of the Deutsche Forschungsgemeinschaft (DFG) through a research infrastructure grant INST 275/257-1 FUGG, CRC 1375 NOA (Project B2), SPP2244 (Project TU149/13-1) as well as DFG grant TU149/16-1. This project has also received funding from the joint European Union's Horizon 2020 and DFG research and innovation programme FLAG-ERA under grant TU149/9-1.
 (+) D.B.,  I.P.\ and Z.G.\ contributed equally to this work. ($\ddagger$) T.L. now at Karlsruhe Institute of Technology, Laboratory for Electron Microscopy, 76131 Karlsruhe, Germany.


\end{document}